\documentclass[twocolumn,
 amsmath,amssymb,
]{revtex4}

\usepackage{soul}

\usepackage{dcolumn}   
\usepackage{bbm}        
\usepackage{amssymb}
\usepackage{dsfont}
\usepackage{caption}
\usepackage{xcolor}
\usepackage[normalem]{ulem}

\usepackage{mathtools,amssymb,amsthm,amsmath}

\usepackage{graphicx}
\usepackage{dcolumn}
\usepackage{bm}

\makeatletter
\let\old@mathcal=\mathcal
\usepackage{dutchcal}
\let\old@dutchcal=\mathcal
\renewcommand{\mathcal}[1]{%
 \if A#1
     \old@mathcal{A}
  \fi
  \if a#1
     \old@dutchcal{a}
  \fi
   \if D#1
     \old@mathcal{D}
  \fi
  \if d#1
     \old@dutchcal{d}
  \fi
   \if I#1
     \old@mathcal{I}
  \fi
  \if i#1
     \old@dutchcal{i}  
  \fi
    \if J#1
     \old@mathcal{J}
  \fi
  \if j#1
     \old@dutchcal{j}  
  \fi
     \if G#1
     \old@mathcal{G}
  \fi
  \if g#1
     \old@dutchcal{g}  
  \fi
 \if N#1
     \old@mathcal{N}
  \fi
  \if n#1
     \old@dutchcal{n}
  \fi
   \if P#1
     \old@mathcal{P}
  \fi
  \if p#1
     \old@dutchcal{p}
  \fi
    \if V#1
     \old@mathcal{V}
  \fi
  \if v#1
     \old@dutchcal{v}
  \fi
   \if R#1
     \old@mathcal{R}
  \fi
  \if r#1
     \old@dutchcal{r}
  \fi
  \if Y#1
     \old@mathcal{Y}
  \fi
  \if y#1
     \old@dutchcal{y}
  \fi
  \if X#1
     \old@mathcal{X}
  \fi
  \if x#1
     \old@dutchcal{x}
  \fi
}

\makeatother

\newcommand{\grafe}[1]{\left\{ #1 \right\}}
\newcommand{\tonde}[1]{\left( #1 \right)}
\newcommand{\quadre}[1]{\left[ #1 \right]}

\begin{document}

\preprint{APS/123-QED}

\title{Generalized Lotka-Volterra equations with random, non-reciprocal interactions:\\
the {typical} number of equilibria}

\author{Valentina Ros}
 \affiliation{Universit\'{e} Paris-Saclay, CNRS, LPTMS, 91405, Orsay, France}
\author{Felix Roy}
\author{Giulio Biroli}
\affiliation{Laboratoire de Physique de l’Ecole Normale Sup\'{e}rieure, ENS, Universit\'{e} PSL,
CNRS, Sorbonne Universit\'{e}, Universit\'{e} de Paris, F-75005 Paris, France}
\author{Guy Bunin}
\author{Ari M. Turner}
\affiliation{Department of Physics, Technion-Israel Institute of Technology, Haifa 32000, Israel}

\begin{abstract}
We compute the typical number of equilibria of the Generalized Lotka-Volterra equations describing species-rich ecosystems with random, non-reciprocal interactions using the replicated Kac-Rice method. We characterize the multiple-equilibria phase by determining the average abundance and similarity between equilibria as a function of their diversity (i.e. of the number of coexisting species) and of the variability of the interactions. We show that linearly unstable equilibria are dominant, and that the typical number of equilibria differs with respect to the average number. 
\end{abstract}

\maketitle

Systems of many degrees of freedom with heterogeneous and \emph{non-reciprocal} (asymmetric) interactions emerge naturally when modelling neural networks \cite{amari1972characteristics, parisi1986asymmetric, hertz1986memory, CrisantiSompolinsky87, DerridaGardnerZippelius,cessac1995increase,aguirre2022satisfiability,stubenrauch2022phase}, natural ecosystems \cite{bascompte2006asymmetric, loreau2013biodiversity, allesina2012stability,galla2006random}, economic networks or agents playing games \cite{ choi2013financial,alfarano2005estimation, mcavoy2015asymmetric, bayer2021best}. The dynamics of these systems are characterized by a large number of attractors such as equilibria, limit cycles and chaotic attractors. Systems admitting an energy landscape, as it is the case for symmetric interactions, only display equilibria, which are the stationary points of the landscape. A rugged landscape is central in the theory of glassy systems, since local minima are associated to metastable states; as a consequence,  in-depth investigations and refined tools for counting and classifying local minima of highly non-convex landscapes have been developed extensively in the context of glassy physics  \cite{monasson1995structural, franz1995recipes,cavagna1997structure,PUZbook}.
Most of these studies focused on systems admitting an energy landscape, though.
Recently, the interest in non-conservative systems (devoid of an energy landscape) has grown substantially and pioneering works have shown that such systems can also display many equilibria \cite{berg1999entropy,schreckenberg1992attractors,fyodorov2016topology,FyodorovNonlinearAnalogue, fedeli2021nonlinearity}. Developing a general theory in order to count them and to investigate  their  \emph{stability} 
is a challenging goal, with potentially relevant implications for understanding the dynamics.

 \begin{figure}[t]
\includegraphics[width=0.51\textwidth]{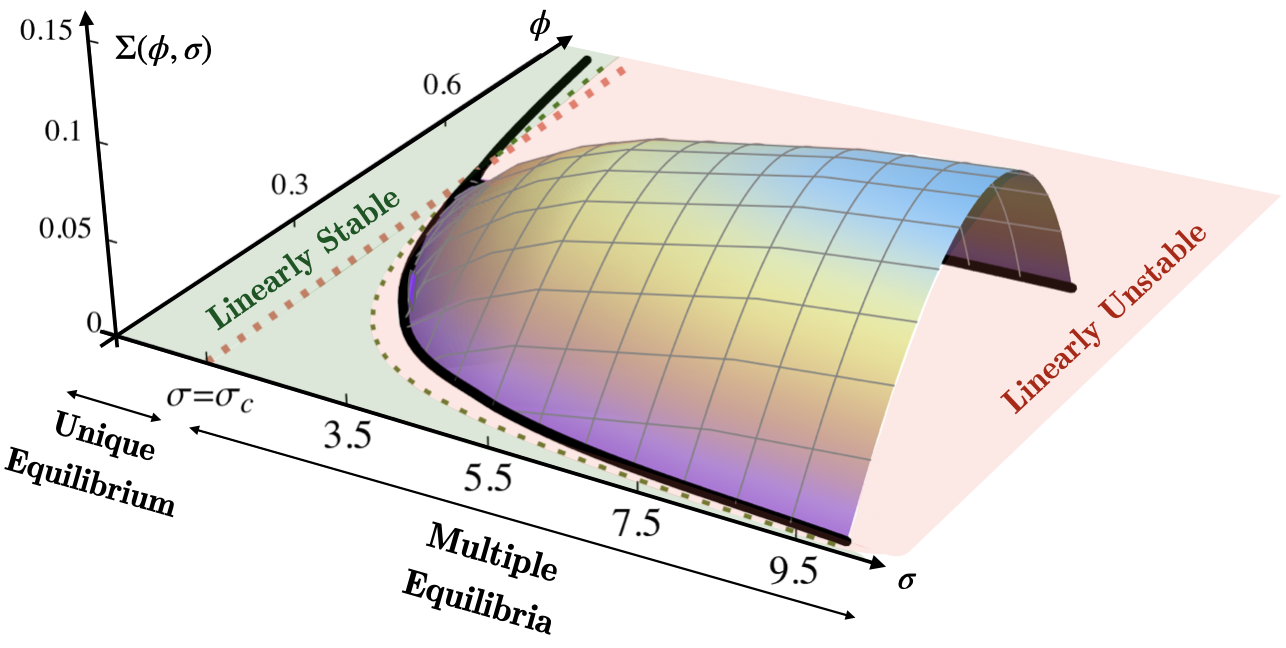}
\caption{Quenched complexity  $\Sigma(\phi, \sigma)$ of uninvadable equilibria for uncorrelated interactions ($\gamma=0$). Black lines correspond to vanishing complexity; the green dotted line to the diversity $\phi_{\rm May}(\sigma)$ above which equilibria are linearly unstable (red area); the orange dotted line to the transition between the unique   ($\sigma<\sigma_c$) and the multiple ($\sigma>\sigma_c$) equilibria phases.}\label{fig:PhaseDiagBunin}
\end{figure}

Here we address this problem for a prototypical non-conservative dynamical system, the random Generalized Lotka-Volterra model (rGLV), which describes the dynamics of population sizes of multiple species with pairwise interactions between them. The rGLV equations are used extensively in theoretical ecology to describe well-mixed ecosystems \cite{goel1971volterra,GuyCavity,jansen2003complexity,barbier2018generic, galla2018dynamically, hu2021emergent}, and they are related to models used in evolutionary game theory and in economic theory 
\cite{moran2019may,diederich1989replicators,galla2013complex,garnier2021new}.
They are known to admit a multiple equilibria phase when the {variability} of the random interactions is strong enough \cite{GuyCavity,LVMarginality,lischke2017finding,AltieriRoyTemperature}, an interesting feature for theoretical ecology \cite{BarbierFingerprints,petraitis2013multiple}.
Our main result is a full characterization of multiple equilibria in terms of average abundance, diversity and stability as summarized in the phase-portrait of Fig. \ref{fig:PhaseDiagBunin}. There is a general expectation that the vast majority (if not all) of the equilibria are linearly unstable when the interactions are asymmetric \cite{opper1992phase, FyodorovNonlinearAnalogue}; our analysis confirms this surmise, which directly implies a complex dynamical behavior, as the system can never settle in a fixed point, even at long times.
In order to properly count the \emph{typical} number of equilibria, we combine
 random matrix theory with standard tools in the theory of glasses. We thus go beyond the previous analysis performed for systems with asymmetric interactions \cite{OpperAnnealedComplexity,berg1999entropy,schreckenberg1992attractors,fyodorov2016topology,FyodorovNonlinearAnalogue}, which focused on the \emph{average} number of equilibria. The latter is in fact much larger than the former and not representative of the typical behavior of the rGLV model, as we shall show below (and as it happens in many other disordered and glassy systems).

The rGLV equations determine the dynamics of a pool of $S \gg 1$ species. They read 
\begin{equation}\label{eq:SystemEqs}
\frac{d{N}_i(t)}{dt}=N_i(t) F_i(\vec{N}), 
\end{equation}
where $N_i(t) \geq 0$ is the {abundance} of species $i$ at time $t$. The vector $\vec{F}$ represents the {growth rates} or forces:
\begin{equation}\label{eq:GradDef}
F_i(\vec{N})=\kappa_i -  N_i -\frac{\mu}{S}  \sum_{j =1}^S  N_j -\frac{\sigma}{\sqrt{S}}\sum_{j =1}^S   a_{ij} N_j.
\end{equation}
Here $\kappa_i$ are the carrying capacities,  $\mu, \sigma$ are the average interaction strength and the variability, and $a_{ij}$ are  components of a random matrix encoding the fluctuations in the interactions between the different species \footnote{The self-interactions are often absorbed in the carrying capacities ($a_{ii}=0$). In order for the calculations with random matrices to work easily, we consider $a_{ii} \neq 0$ with the same statistics as for the off-diagonal $a_{ij}$. This choice does not affect any large-$S$ result discussed in the work. }. To describe interactions where $a_{ij}$ and $a_{ji}$ are correlated but not exactly the same, we take them as
two variables with a joint Gaussian distribution defined by covariances:
\begin{equation} 
\langle a_{ij} a_{kl} \rangle=\delta_{i k} \delta_{jl} + \gamma \; \delta_{il} \delta_{jk}, \quad |\gamma|\leq 1
\end{equation}
corresponding to  $\langle a_{ij}^2\rangle=\langle a_{ji}^2\rangle=1$ and
$\langle a_{ij} a_{ji} \rangle=\gamma$. In the extreme case $\gamma=\pm 1$ one obtains perfect correlations $a_{ij}=\pm a_{ji}$, while for $\gamma=0$ the interactions are uncorrelated. We focus on  $\kappa_i=\kappa$, but the calculation can be easily generalized to heterogeneous $\kappa_i$.

\begin{figure}[ht]
  \centering 
    \centering \includegraphics[width=\linewidth]{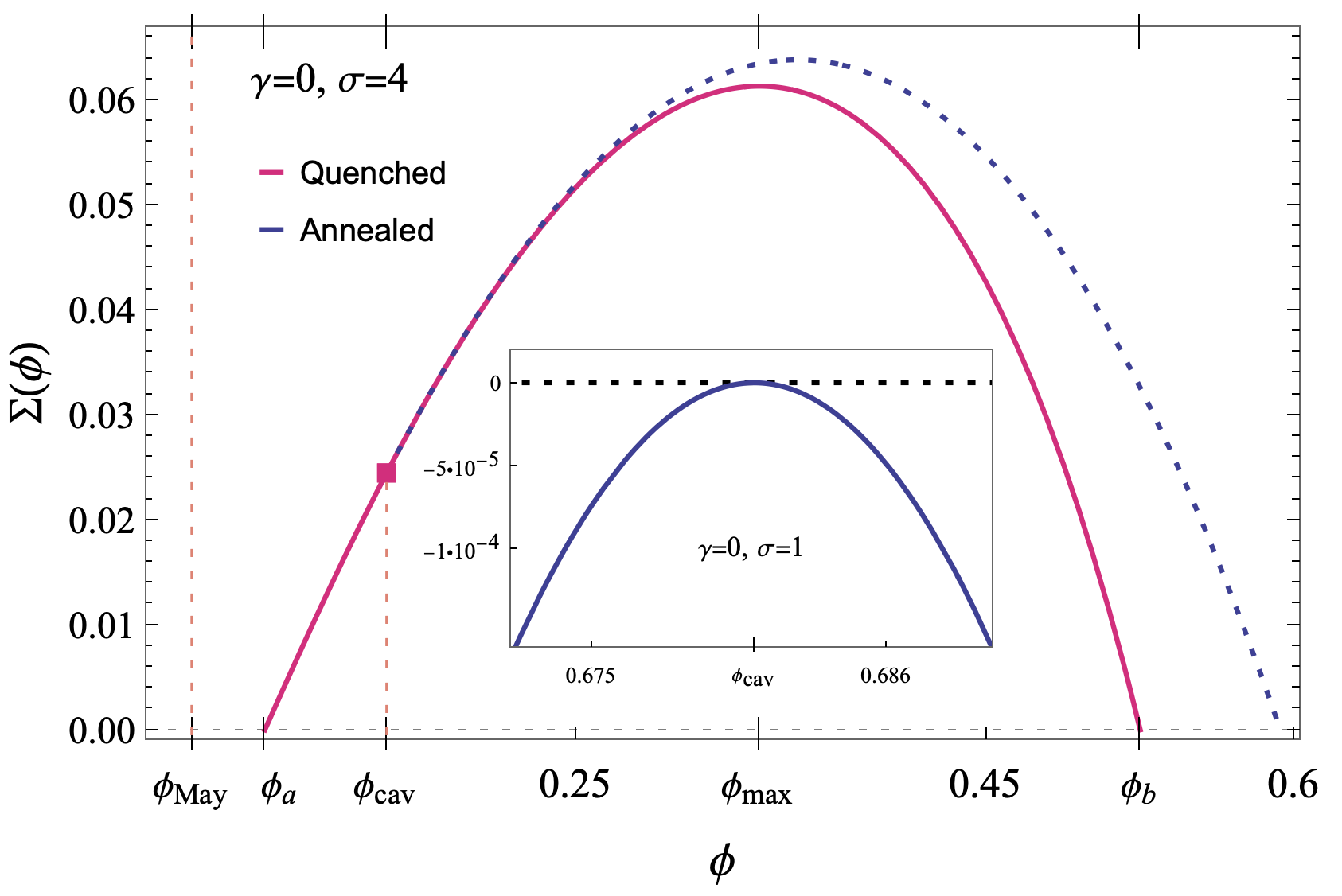}
\caption{Complexity of equilibria as a function of their diversity, for $\gamma=0$. Main panel: Complexity in the multiple equilibria phase (at $\sigma=4$). A difference between quenched (magenta) and annealed (blue) is apparent. All the equilibria are unstable ($\phi>\phi_{\rm May}$). Inset: Annealed complexity in the unique equilibrium phase (at $\sigma=1$),  negative except at
the diversity predicted by the cavity formalism consistent with the existence of a unique equilibrium.}\label{fig:RepresentativePlotComp}
\end{figure}

Equilibria are configurations $\vec{N}^*$ satisfying 
\begin{equation}\label{eq:SystemEqs2}
\frac{d N^*_i}{dt}=N^*_i  F_i(\vec{N}^*)=0 \quad  \forall \;  i, \quad \quad N^*_i \geq 0.
\end{equation}
Numerical simulations and analytical results \cite{rieger1989solvable,RoyDMFT,GuyCavity, galla2018dynamically, opper1992phase,LVMarginality} reveal two distinct regimes for large $S$: 
 a \emph{unique equilibrium} regime in which any arbitrary initialization of the population vector converges to a fixed equilibrium $\vec N^*$ which is globally stable, and a \emph{multiple equilibria} regime. The transition 
 between the two regimes takes place 
at $\sigma_c=\sqrt{2}(1+ \gamma)^{-1}$  \cite{rieger1989solvable}. 
Characterizing the \emph{multiple equilibria} phase when $-1<\gamma<1$ is still an open challenge as mappings to physical systems work only for $\gamma=1$  \cite{diederich1989replicators,biscari1995replica,GuyCavity,LVMarginality,AltieriRoyTemperature, marcus2022local} and $\gamma=-1$ \cite{pearce2019stabilization}.
In the former case the problem is conservative and the force is obtained as the derivative of an energy, $F_i(\vec{N})= - \partial_{i} L(\vec{N})$ with $L(\vec{N})=  \sum_{i=1}^S N_i[\frac{N_i}{2}-\kappa_i + \frac{\mu}{2} \sum_{j=1}^S  N_j + \frac{\sigma}{2 \sqrt{S}} \sum_{j=1}^S a_{ij} N_j]$. Stable equilibria are identified with metastable states (local minima of the energy). Spin-glass techniques \cite{LVMarginality, AltieriRoyTemperature} can be used to show that there exist exponentially many (in $S$) metastable states, the relevant ones being marginally stable, which makes the system critical \cite{muller2015marginal} and hence very fragile to non-conservative perturbations \cite{spitzner1989freezing,opper1992phase, iori1997stability, cugliandolo1997glassy}. This formalism requires the existence of an energy landscape. When $-1<\gamma< 1$,
Dynamical Mean Field Theory \cite{RoyDMFT} has provided information on the dynamics but not directly on the equilibria. Here we tackle this challenge 
by the Kac-Rice formalism \cite{fyodorov2013high,fyodorov2005counting,fyodorov2004complexity, auffinger2013random}. To 
study the \emph{typical} number of equilibria for $\gamma \neq 1$ we make use of the so called quenched
Kac-Rice formalism introduced in~\cite{ros2019complex}.

There are many equilibria solving \eqref{eq:SystemEqs2}, that differ by which species are present. We classify their typical number as a function of their \emph{diversity}: each equilibrium $\vec{N}^*$ has a certain number of absent species ($N_i^*=0$), and a number $s(\vec{N}^*) $ of present species ($N_i^*>0$). The diversity is defined as $\phi(\vec{N}^*)=s(\vec{N}^*)/S \in [0,1]$. This quantity is a central property in ecology, which also sets the stability of the equilibria \cite{may1972will}, as we recall below. 
Our counting of equilibria at varying $\phi$ is also motivated by the fact that it is not known a priori which equilibria will affect the systems dynamics (and how), at variance with equilibrium frameworks where the relevant equilibria are marginally stable minima, usually the more numerous ones (see however \cite{folena2020rethinking}). Therefore, determining the \emph{range} of diversities where equilibria are present is crucial.
We focus on \emph{uninvadable} equilibria,
such that $F_i(\vec{N}^*)<0$ for any $i$ such that $N_i^*=0$ (notice that similar constraints appear naturally in constraint satisfaction problems, too \cite{franz2016simplest}). These equilibria are relevant as they are stable with respect to small positive fluctuations in the abundance of the absent species. 
The total number $\mathcal{N}_{S}(\phi)$ of uninvadable equilibria with diversity $\phi$ scales exponentially with $S$ \cite{Fried2017}. As known from glassy physics,  $\mathcal{N}_{S}(\phi)$ is  a random variable which in general does not concentrate around its average (it is not self-averaging). In this case the typical number is obtained by focusing on  
the large-$S$ limit of its logarithm, which does concentrate around a deterministic value $\Sigma(\phi)$: 
\begin{equation}\label{eq:QuenchedComp}
 \lim_{S \to \infty} \frac{  \log \quadre{ \mathcal{N}_{S}(\phi)} }{S}  =  \lim_{S \to \infty} \frac{\left \langle  \log \quadre{ \mathcal{N}_{S}(\phi)} \right \rangle}{S} \equiv \Sigma(\phi ).
\end{equation}
 $\Sigma(\phi)$ governs the exponential scaling of the \emph{typical} value of $\mathcal{N}_{S}(\phi)$: borrowing the terminology from  glassy physics, we refer to it as the \emph{quenched complexity}. The computation of the average of the logarithm is done via the replica trick:
\begin{equation}\label{eq:RepTrick}
 \left \langle \log { \mathcal{N}_{S}(\phi )} \right \rangle = \lim_{n \to 0} \frac{\log \left \langle { \mathcal{N}^n_{S}(\phi )} \right \rangle}{n}.
\end{equation}
When evaluated at $n=1$ the right hand of side of eq. (\ref{eq:RepTrick}) gives the \emph{annealed complexity} associated with the average number of equilibria  \cite{FyodorovNonlinearAnalogue,galla2007two,Toboul, arous2021counting}:
$\Sigma^{(A)}(\phi )  \equiv \lim_{S \to \infty} \frac{1}{S} \log \left \langle { \mathcal{N}_{S}(\phi )} \right \rangle$. When $\mathcal{N}_{S}(\phi)$ is not self averaging, $\Sigma^{(A)}>\Sigma$: the average of $\mathcal{N}_{S}(\phi)$ is dominated by exponentially rare ecosystems displaying an unusually large number of equilibria. It is therefore much larger than the typical value, which captures the properties of the ecosystems occurring with probability that is not suppressed exponentially in $S$.

The main steps of the replicated Kac-Rice computation are explained in the SI. The value of $\left \langle { \mathcal{N}^n_{S}(\phi )} \right \rangle$ can be determined by introducing $n$ copies of the ecosystem and by  
finding the probability that any $n$ given vectors $\vec N^a$, $a=1,\cdots, n$  satisfy Eq. \eqref{eq:SystemEqs2} simultaneously, together with the uninvadability condition. This is a function of order parameters measuring properties of the equilibria, like the  amount of correlation between them. The number of equilibria is dominated (according to a large deviation principle) by specific values of these order parameters. 
 The order parameters are the first two empirical moments of the vectors $\vec N^a$ and $\vec F^a$, i.e. the $2n$ quantities:
\begin{equation}\label{eq:OP}
 m_a \equiv \lim_{S \to \infty} \frac{\sum_{i=1}^S N^a_i}{S}, \quad  p_a \equiv \lim_{S \to \infty} \frac{\sum_{i=1}^S F^a_i}{S} 
\end{equation}
as well as the $n(n+1)$ +$n(n-1)$ correlations (or \emph{overlaps}):
\begin{equation}
\begin{split}
  q_{ab}\equiv \lim_{S \to \infty}& \frac{\vec N_a \cdot \vec N_b}{S}, \;   \xi_{ab}\equiv \lim_{S \to \infty} \frac{\vec F_a \cdot \vec F_b}{S}, \\  & z_{ab}\equiv \lim_{S \to \infty} \frac{\vec N_a \cdot \vec F_b}{S}
  \end{split}
  \end{equation}
where $z_{aa}=0$ follows from \eqref{eq:SystemEqs2}. These order parameters encode the correlations in the location of the different fixed points in configuration space, which emerge because all the fixed points arise from the same interactions between the species. We consider a symmetric ansatz for the order parameters, i.e. $ m_a=m,  q_{ab}=\delta_{ab} q_1 + (1-\delta_{ab})q_0,
     p_a=p,  \xi_{ab}=\delta_{ab} \xi_1 + (1-\delta_{ab})\xi_0 z_{ab}=(1-\delta_{ab})z,$
     which is the simplest approximation that takes such correlations into account.
Under this assumption, the moments can be written as an integral over all possible values of the order parameters:
\begin{equation}\label{eq:bSP-rs2}
\left \langle \mathcal{N}^n_S(\phi) \right \rangle = \int d {\bf x} \; e^{S \, n \, \bar{\mathcal{A}}({\bf x}; \phi) + o(n S)},
\end{equation}
with ${\bf x}=(m,p,q_1, q_0, \xi_1, \xi_0, z)$, see the SI for details of the calculation of $\bar{\mathcal{A}}$ and for its explicit expression.
 The large deviation principle then implies that asymptotically
\begin{equation}\label{eq:Action}
\Sigma(\phi)= \bar{\mathcal{A}}({\bf x}^\star; \phi),
\end{equation}
where ${\bf x}^\star$ is the solution of the saddle-point equations  $\frac{\delta \bar{\mathcal{A}}({\bf x} ; \phi)}{\delta {\bf x}} \Big|_{{\bf x}^\star}=0$.
This results in self-consistent equations for the typical properties of equilibria at fixed $\phi$, such as their typical  average abundance $m^*$ or the typical similarity between two equilibria $q_0^*$. 

The Kac-Rice computation allows us to determine the linear stability of the equilibria at each given $\phi$ with respect to perturbations  $N^*_i \to N^*_i + \delta N^*_i$ of the populations of coexisting species. This depends on the spectral properties of the matrix:
\begin{equation}\label{eq:Hessian}
H_{ij}(\vec{N}^*)=  \tonde{\frac{\partial F_i (\vec{N}^*)}{dN_j} }_{i, j: {N}_i^*, N^*_j >0}.
\end{equation}
For stable equilibria all the eigenvalues of \eqref{eq:Hessian} have negative real part. The asymmetry of the  matrix $a_{ij}$ implies that \eqref{eq:Hessian} are themselves asymmetric random matrices \cite{CrisantiSommersStein}. 
The typical eigenvalue density (neglecting possible isolated eigenvalues) of $H_{ij}$  depends on  $\vec{N}^*$ only through its diversity $\phi$. For 
\begin{equation}\label{eq:MayBound}
\phi < \phi_{\rm May}= \frac{1}{\sigma^2 (1+ \gamma)^2}.
\end{equation}
the density has support on the negative real sector; 
therefore a typical equilibrium with $\phi<\phi_{\rm May}$ (if it exists) is stable. At $\phi= \phi_{\rm May}$, the support of the eigenvalue density touches zero and the corresponding equilibrium is marginally stable; for larger $\phi$ the equilibrium is unstable. The criterion \eqref{eq:MayBound} for linear stability is related to that identified by May in \cite{may1972will}, and we henceforth refer to it as the \emph{May stability bound}. 
More details on the Kac-Rice computation, with a thorough discussion of the structure of the equations and their resolution, are given in~\cite{PaperLungo}.

We now present our main results, focusing on the case of {uncorrelated} interactions $\gamma=0$ and setting $\kappa=1$. We find that although the saddle point values ${\bf x}^*$ depend explicitly on $\mu$, the complexity at fixed diversity does not, allowing us to discuss the behavior of $\Sigma(\phi)$ as a function of $\sigma$ only. As shown in Fig. \ref{fig:PhaseDiagBunin},  when $\sigma>\sigma_c$ there is a range of diversities $\phi \in [\phi_a(\sigma), \phi_b(\sigma)]$ for which $\Sigma(\phi)>0$  (a negative annealed $\Sigma(\phi)$  signifies that no equilibria exist typically \cite{auffinger2013random}). The rGLV equations thus admit an exponentially large number of uninvadable equilibria with a continuous
distribution of diversities. All the equilibria are unstable, as their diversity exceeds the May stability bound, Eq.~\eqref{eq:MayBound}. 
In Fig.~\ref{fig:RepresentativePlotComp} we show a cut at fixed $\sigma$ of the plot of Fig. \ref{fig:PhaseDiagBunin}. 
In addition to the quenched complexity we show the annealed one for comparison. We find that the complexity and the diversity $\phi_{\rm max}$
associated to the typical, i.e. most numerous equilibria at the given $\sigma$ are 
overestimated by the annealed calculation. Annealed and quenched complexity only coincide for small $\phi$.  The point $\phi_{\rm cav}$ where they begin to deviate from one another turns out to coincide with the value of diversity predicted by the cavity method discussed in Refs.~\cite{opper1992phase, GuyCavity, galla2018dynamically}.  The cavity method assumes the existence of a unique stable equilibrium and allows one to characterizes its abundance $m$ and overlap $q_1$, by imposing consistency relations between the properties of the system with $S+1$ and $S$ species. The above result shows that despite being only approximate for $\sigma>\sigma_c$, this method still captures the properties of a given family of equilibria, even though they are exponentially rare with respect to the typical ones at $\phi_{\rm max}$.

\begin{figure}[ht]
  \centering  \includegraphics[width=.97\linewidth]{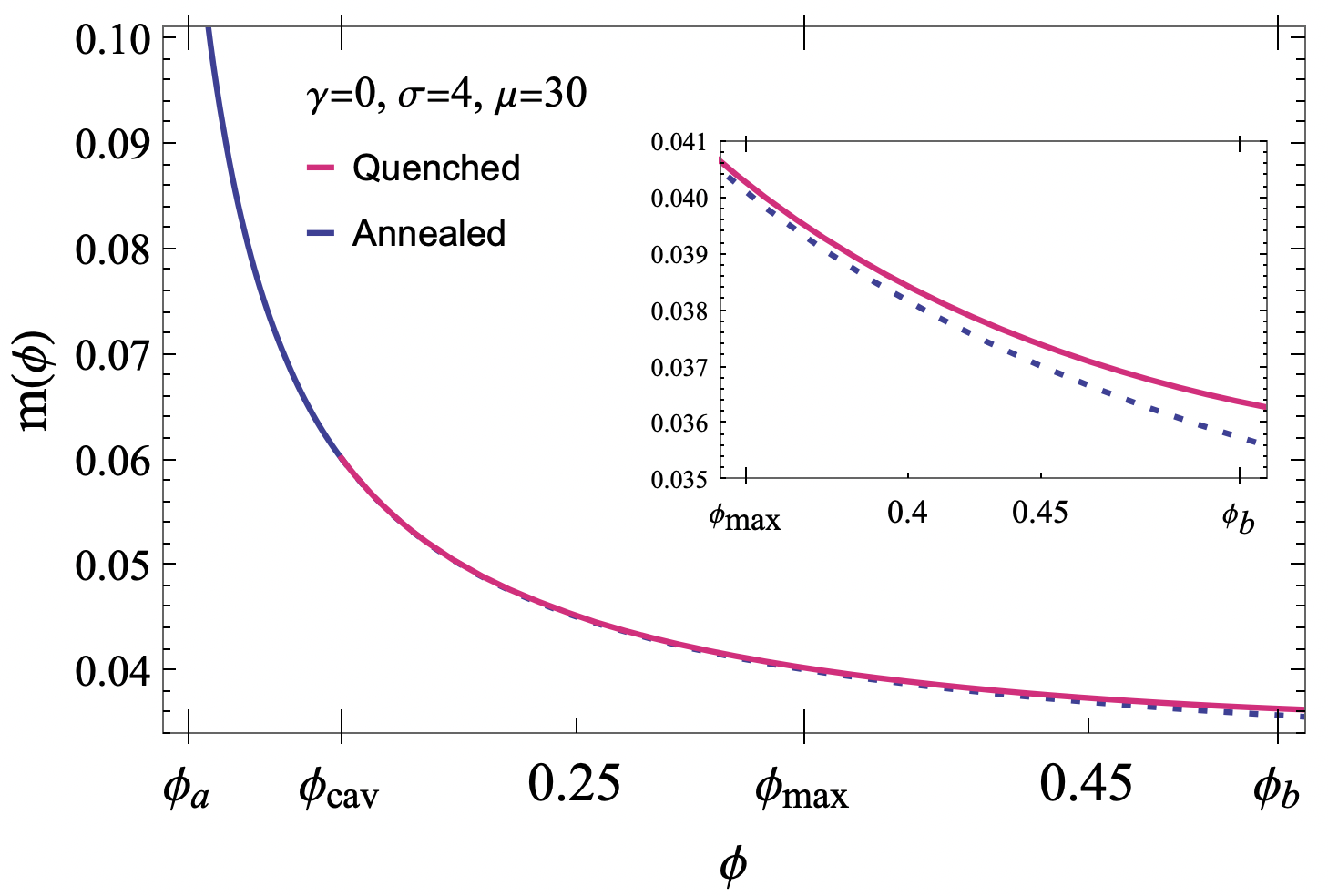}\par
\caption{Typical averaged population size as a function of  diversity $\phi$ for $\sigma=4$ and $\mu=30$, in the annealed (blue) and quenched (magenta) calculation. More diverse equilibria have a smaller averaged population size $m$, which for $\phi>\phi_{\rm cav}$ is underestimated by the annealed approximation. The inset is a zoomed plot. }\label{fig:OrderParPhi}
\end{figure}

\begin{figure}[ht]
  \centering  \includegraphics[width=.97\linewidth]{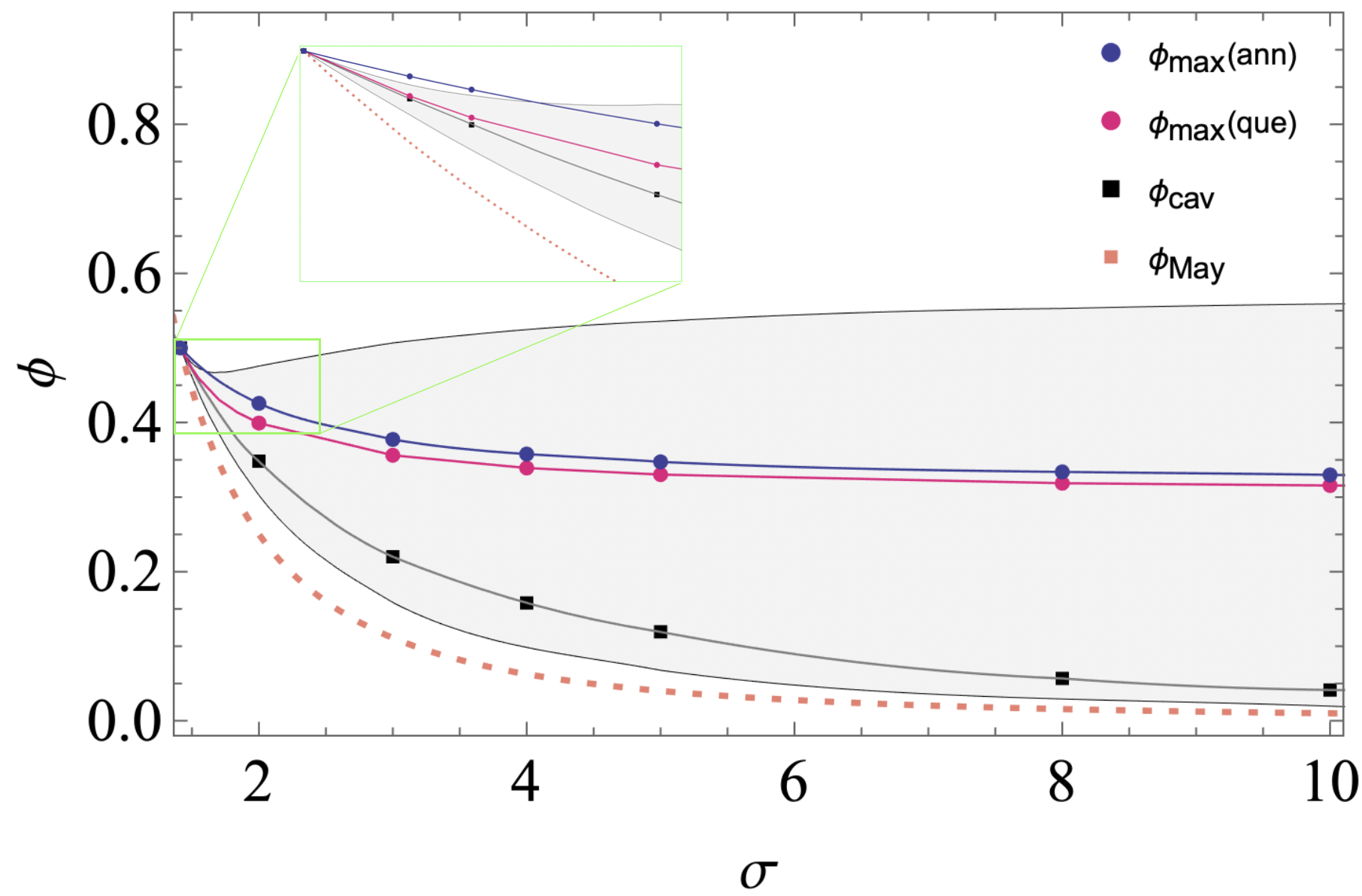} 
\caption{ Diversity vs. variability diagram. The range of possible diversities is indicated by the grey region. Curves of maximal complexity are shown in magenta (quenched) and blue (annealed). The black squares give $\phi_{\rm cav}$. The orange dashed line corresponds to $\phi_{\rm May}$ above which all equilibria are linearly unstable. \emph{Inset. } Zoom in the vicinity of $\sigma_c=\sqrt{2}$. }\label{fig:DivSigmaPreliminary}
\end{figure}

We have studied how the properties of equilibria change as $\phi$ is increased. Fig.~\ref{fig:OrderParPhi} shows that imposing a larger diversity leads to less populated (lower average abundance $m^*$) equilibria. Similarly, it leads to less correlated (lower overlap $q_0^*$) equilibria. 
Fig.~\ref{fig:DivSigmaPreliminary} shows the $\sigma$-dependence of the special values of $\phi$ discussed above (it corresponds to Fig.1 seen from the top). The grey area is the support of the quenched complexity, which increases with $\sigma$. When $\sigma \to \sigma_c^+$ all the special values of $\phi$ merge together and reach 
$\phi_{\rm May}$. Correspondingly the complexity vanishes.

Just above $\sigma_c$, where the complexity goes to zero, the quenched and annealed calculations have great discrepancies, see the inset of Fig. \ref{fig:DivSigmaPreliminary}, probably due to the larger correlation between equilibria. In fact, 
the average number of equilibria (annealed calculation) is dominated by equilibria having a diversity $\phi_{\rm max}^{\rm ann}$ for which \emph{typically} there are no equilibria, i.e. the quenched complexity vanishes. {This feature had already been identified in Ref.~\cite{garnier2021new} for a slightly different model arising in the context of portfolio optimization (and describing, in its ecological interpretation, species competing for a single common resource). }

For larger $\sigma$ the cavity approximation underestimates more strongly the diversity (and thus the instability) with respect to that of typical equilibria at $\phi_{\rm max}$. 
For $\sigma< \sigma_c$, the complexity (annealed and quenched) is non-negative only at $\phi=\phi_{\rm cav}$, 
which now correctly describes the diversity of the system as there is a unique equilibrium \cite{clenet2022equilibrium}.
The analysis of the multiple equilibria also allows us to characterize thoroughly the transition to  an additional phase, the unbounded phase, where some abundances diverge  as a function of time, see the SI.

Finally, let us focus on the properties of the transition to the unique equilibrium phase at $\sigma_c$. Following the terminology introduced in \cite{fyodorov2016topology}, this is \emph{a trivialization transition} and corresponds to the point at which the total quenched complexity $\Sigma_{\rm tot}= \Sigma(\phi_{\rm Max})$ first vanishes. The way in which $\Sigma_{\rm tot}$
vanishes for $\sigma \to \sigma_c$ has been the focus of several works. It has been studied in models with a quadratic  single-species confinement potential within an annealed calculation \cite{arous2021landscape,lacroix2022counting}. Importantly, it has also been conjectured to be connected to the emergence of chaos and of a finite Lyapunov exponent \cite{Toboul}.
For the rGLV model at $\gamma=0$ we find that the complexity grows quadratically with $\sigma$ when entering the multiple equilibria phase, $\Sigma_{\rm tot} \sim (\sigma-\sigma_c)^2$ as in \cite{arous2021landscape,lacroix2022counting}. 
As found in models of recurrent neural networks, the emergence of a non-zero complexity is concomitant with the emergence of a complex dynamical behavior, including chaos and aging \cite{RoyDMFT}. We notice that the annealed approximation locates correctly the trivialization transition in this case, and also captures the quadratic increase but with a different prefactor. We do not expect this quadratic behavior to be general, unless the total complexity in the vicinity of $\sigma_c$ is captured by the annealed framework. If this is not the case, our calculation suggests that one should find a different power law for $\gamma \neq 0$ (see~\cite{PaperLungo} and the SI for more details).

In summary, we have characterized the multiple-equilibria phase of the rGLV equations by computing  explicitly the complexity of uninvadable equilibria. On a technical ground our approach, giving access to the quenched complexity, has allowed us to assess when and to what degree the annealed calculation is precise: we have found a transition at the value of diversity $\phi_{\rm cav}$, below which the annealed calculation is exact and above which the quenched calculation gives a quantitatively different result;  the latter regime always includes the maximum of the complexity, which corresponds to the typical equilibria.

We performed the calculation assuming a symmetry of the order parameters with respect to permutations of replicas: we are thus restricting the region of parameter space where to look for solutions of the self-consistent equations obtained from the variation of \eqref{eq:Action}. For $\gamma=1$ it is know that the symmetric assumption is an approximation, as \eqref{eq:Action} is optimized by parameters that break the symmetry between the replicas. Verifying that Replica Symmetry Breaking (RSB) is not needed for generic $\gamma$ is a challenge that we leave for further studies.

Our calculations show that for non-reciprocal uncorrelated interactions all the uninvadable equilibria are linearly unstable. This marks a difference with respect to the symmetric case, where marginally stable equilibria are present and correspondingly the dynamics is glassy. 
With unstable equilibria, a chaotic dynamics is expected in presence of migration~\cite{cessac1995increase} and signatures of it emerge in theoretical models~\cite{KesslerShnerb2015} and even in controlled experiments~\cite{beninca2008chaos}. Similarly to the case of landscape studies which were instrumental to understand glassy dynamics in terms of local minima and metastable states, it would be very interesting to connect the properties of 
these unstable equilibria (more generally, of heteroclinic networks formed by them~\cite{o2021intrinsic})
to the dynamical behavior. 
We envisage that \emph{invadable} equilibria also play a role in the dynamics~\cite{de2022aging}, and the calculation of their complexity is ongoing, as well as the generalization to inhomogeneous carrying capacities $\kappa_i$~\cite{bouchbinder2021low, arous2021landscape, mergny2021stability}.

\subsection*{Acknowledgements}
We thank D. Fisher for discussions on this topic.
 VR also thanks P.~Urbani for hints on the calculation of the quenched complexity, and B.~Lacroix-A-Chez-Toine, J.~Berg, J.~Krug and M.~Mungan for suggestions of references. VR acknowledges funding by the ``Investissements d’Avenir” LabEx PALM (ANR-10-LABX-0039-PALM).

\bibliographystyle{ieeetr}
\bibliography{bibLV2}

\onecolumngrid
\appendix

\vspace{2 cm}

{\Large \centering \bf Supplemental Material} \\

We report in the following the main steps to obtain the quantity $\bar{\mathcal{A}}({\bf x}; \phi)$ appearing in Eq.~\eqref{eq:bSP-rs2} in the main text. Moreover, we discuss additional results on the unbounded phase and on the vanishing of the total complexity, which are mentioned in the main text. For a more detailed exposition of the formalism underlying this calculation, we refer the reader to Ref. \cite{PaperLungo}. \\

\textbf{The Kac-Rice formula for the moments. }
The Kac-Rice formalism is a framework that allows one to characterize the number of solutions of dynamical equations containing randomness: in particular, given that the number of solutions is itself a random variable, the formalism gives a recipe to determine the moments of this random variable. For an introduction to the formalism and to its application to the high-dimensional setting, see \cite{fyodorov2013high, ros2022high} and references therein. This formalism provides us with an expression for the moments of the number of equilibria at fixed diversity, denoted with $\mathcal{N}_{S}(\phi)$ in the main text. To compute the $n$-th moment of this random variable, we need to introduce $n$ different configurations $\vec{N}^{a}$ of the ecosystem (with $a=1, \cdots n$), which we refer to as \emph{replicas.} Each $\vec{N}^{a}$ represents a realization of the ecosystem at fixed values of the rand interaction terms $a_{ij}$. We let ${\bf N}= (\vec{N}^{1}, \cdots, \vec{N}^{n})$ denote the concatenation of configurations of all replicas. In each configuration $\vec{N}^{a}$, some species will be present ($N_i^a>0$) while some others will be absent ($N_i^a=0$). We let  $I_a = I(\vec{N}^{a})$ be the index set collecting the indices of the species that are present in the configuration $\vec{N}^{a}$. Since we are interested in counting the equilibria having fixed diversity $\phi$, we enforce that  $|I_a|=S \phi$ for all $a$. We introduce the vectors of growth rates or forces $\vec{F}^{a}= \vec{F}(\vec{N}^{a})$ and ${\bf F}({\bf N})=(\vec{F}^{1}, \cdots, \vec{F}^{n})$. Let ${\bf f}$ denote the value taken by this random vector, and $\mathcal{P}_{{\bf N}} \tonde{ {\bf f}}$ the joint distribution of the $S$-dimensional vectors $\vec{F}^{a}$ evaluated at $\vec{f}^{a}$,
\begin{equation}
\mathcal{P}^{(n)}_{{\bf N}} \tonde{ {\bf f}}=\int \prod_{i,j=1}^S \,d a_{ij} \mathbb{P} (\grafe{a_{ij}}_{ij}) \, \delta \tonde{{\bf F}({\bf N})- {\bf f}}.
\end{equation}
We also introduce the following conditional expectation value:
\begin{equation}\label{eq:ExpectationJointDet}
 \mathcal{D}^{(n)}_{{\bf N}} \tonde{ {\bf f}}= \left \langle \tonde{\prod_{a=1}^n  \left| \text{det} \tonde{\frac{\delta F^{a}_i}{dN_j^{a}}}_{i, j \in I_a}\right| } \; \; \Big| \; {\bf  F}({\bf N})= {\bf f} \right \rangle.
\end{equation}
The latter is the expectation of the product of the absolute values of $n$ determinants of the $S \phi \times S \phi$ matrices of derivatives of the components of ${\bf F}$, conditioned to ${\bf F}$ itself taking value ${\bf f}$.
The Kac-Rice formula for the $n$-th moment of the number  $\mathcal{N}_{S}(\phi)$ of uninvadable equilibria reads:
\begin{equation}\label{eq:KRmoments}
\begin{split}
\left \langle \mathcal{N}^n(\phi) \right \rangle= &\sum_{\begin{subarray}{c}  I_1\\  |I_1|= S \phi  \end{subarray}} \cdots  \sum_{\begin{subarray}{c}  I_n\\  |I_n|= S \phi  \end{subarray}}  \prod_{a=1}^n  \int d\vec{N}^{a}\, d \vec{f}^{a}  \prod_{i \in I_a} \theta (N_i^{a}) \, \delta( f_i^{a}) \prod_{i \notin I_a} \delta( N_i^{a}) \theta( -f_i^{a}) \mathcal{D}^{(n)}_{{\bf N}} \tonde{ {\bf f}} \mathcal{P}^{(n)}_{{\bf N}} \tonde{ {\bf f}}.
\end{split}
\end{equation}
We now briefly summarize how to determine the behaviour of the moments \eqref{eq:KRmoments} for generic values of $n$ to leading exponential order in $S$, and how to extract the quenched (and annealed) complexity from it.\\

\textbf{The order parameters and the complexity. }
By performing the averages over the random interactions $a_{ij}$, one sees that the quantities $\mathcal{D}^{(n)}_{\bf N} \tonde{\bf f}$ and $\mathcal{P}^{(n)}_{\bf N} \tonde {\bf f}$ in \eqref{eq:KRmoments}  depend on the vectors $\vec{N}^{a}$ and $\vec{f}^{a}$ only through their scalar products. For $a,b=1, \cdots, n$ we can therefore introduce a set of \emph{order parameters} defined as follows:
\begin{equation}\label{eq:OrderParameters}
\begin{split}
  S q_{ab}= \vec{ N}^a \cdot \vec{ N}^b,\quad 
   S \xi_{ab}= \vec{ f}^a \cdot \vec{ f}^b,\quad
   S z_{ab}= \vec{ N}^a \cdot \vec{ f}^b,\quad
   S m_a= \vec{ N}^a \cdot \vec{ 1},\quad
   S p_a= \vec{ f}^a \cdot \vec{ 1},
  \end{split}
\end{equation}
where $\vec{ 1}= (1, \cdots, 1)^T$ is an $S$-dimensional vector with all entries equal to one. It follows that the integration over $\vec{N}^a, \vec{f}^a$ in \eqref{eq:KRmoments} can be replaced by an integration over the order parameters, with the appropriate change of variables. The calculation proceeds in a few steps that we briefly summarize. First, the order parameters are introduced in \eqref{eq:KRmoments} by means of the identities:
\begin{equation}
 1= \int dq_{ab}\, \delta \tonde{\frac{\vec{ N}^a \cdot \vec{ N}^b}{S}- q_{ab}} = S \int dq_{ab}  \int \frac{d \hat{q}_{ab}}{2 \pi} e^{i \hat q_{ab} \tonde{\vec{ N}^a \cdot \vec{ N}^b-S q_{ab}}},
\end{equation}
where the auxiliary variables $\hat q_{ab}$ are  \emph{conjugate parameters} (and similarly for the other order parameters in \eqref{eq:OrderParameters}). Then, we make use of the assumption that the order parameters are symmetric with respect to permutations of the replicas, which implies that:
\begin{equation}\label{eq:RSa}
\begin{split}
 q_{ab}= \delta_{ab} q_1+ (1-\delta_{ab}) q_0,\quad
  \xi_{ab}=\delta_{ab} \xi_1+ (1-\delta_{ab}) \xi_0,\quad
  z_{ab}= (1-\delta_{ab}) z, \quad
  m^a= m, \quad
  p^a= p,
  \end{split}
\end{equation} 
and similarly for the conjugate ones. Let then ${\bf x}=(m,p,q_1, q_0, \xi_1, \xi_0)$ denote the collection of all of these order parameters, and $\hat{\bf x}=(\hat m, \hat p, \hat q_1, \hat q_0, \hat \xi_1, \hat \xi_0)$ the collection of the conjugate ones. Performing the integration over $\vec{N}^a, \vec{f}^a$ at fixed values of ${\bf x}, \hat {\bf x}$ and performing an expansion of the resulting expressions for large $S$, one then obtains the following integral representation for the moments:
\begin{equation}\label{eq:bSP-rs}
\left \langle \mathcal{N}^n(\phi) \right \rangle = \int d {\bf x} \,  id \hat{\bf x}\; e^{S  \, \mathcal{A}_n({\bf x} , \hat{\bf x}, \phi) + o(S)},
\end{equation}
where the function $\mathcal{A}_n({\bf x} , \hat{\bf x}, \phi)$ depends only on the order parameters and on the conjugate parameters, as well as on the number $n$ of replicas. Given that $S$ is large, the leading order contribution to the moments can be determined by means of a saddle point approximation, by evaluating $\mathcal{A}_n({\bf x} , \hat{\bf x}, \phi)$ at the stationary point ${\bf x}^*, \hat {\bf x}^*$ which maximizes it. This can be done in principle for arbitrary values of $n$. We recall that the \emph{annealed} complexity is obtained taking the logarithm of~\eqref{eq:bSP-rs} with $n=1$, while the \emph{quenched} complexity is obtained taking the limit $n \to 0$ according to Eq.~\eqref{eq:RepTrick}.
By choosing $n=1$, we obtain:
\begin{equation}\label{eq:ActionRSn1}
\begin{split}
\mathcal{A}_1({\bf x}, \hat{\bf x}, \phi)=\mathcal{p}_1({\bf x})+  \mathcal{d} (\phi)+ \tonde{\hat{q}_1 q_1+\hat{\xi}_1 \xi_1+ \hat m m + \hat p p + \hat \phi \phi}+  \mathcal{J}_1( \hat{\bf x}),
\end{split}
\end{equation}
with 
\begin{equation}
\begin{split}
 \mathcal{p}_1({\bf x})&=-\frac{1}{2 \sigma^2 q_1^2 }\quadre{ (\kappa -\mu  m)^2  \tonde{q_1 -\frac{\gamma\, m^2 }{1+ \gamma} }-2  (\kappa-\mu  m) q_1\tonde{p+\frac{m}{1+\gamma} }+ \xi_1  q_1} -\frac{1}{2} \log (2 \pi \sigma^2 \, q_1) -\frac{1}{2 \sigma^2 (1+\gamma)},
   \end{split}
\end{equation}

\begin{equation}\label{eq:VolumeAnnealed}
\begin{split}
 \mathcal{J}_1(\hat{\bf x})&=    \log \quadre{\frac{1}{2} \sqrt{\frac{\pi}{\hat{\xi}_1}} e^{\frac{\hat{p}^2}{4 \hat{\xi}_1}} \text{Erfc} \tonde{-\frac{\hat{p}}{2 \sqrt{\hat{\xi}_1}}}+\frac{e^{- \hat{\phi}}}{2} \sqrt{\frac{\pi}{\hat{q}_1}} e^{\frac{\hat{m}^2}{4 \hat{q}_1}} \text{Erfc} \tonde{\frac{\hat{m}}{2 \sqrt{\hat{q}_1}}}},
  \end{split}
\end{equation}
and
\begin{equation}\label{eq:DetSmall}
\begin{split}
\mathcal{d}(\phi)=\frac{\phi}{\pi} \int_{-1}^1 dx \int_0^{\sqrt{1-x^2}} dy \log \grafe{\quadre{ \sigma \sqrt{\phi} (1+\gamma)x+1}^2 + \sigma^2 \phi (1-\gamma)^2 y^2}.
   \end{split}
\end{equation}
This double integral can be evaluated explicitly, and one finds: 
\begin{equation}\label{eq:DetSmall22}
\begin{split}
\mathcal{d}(\phi)=\begin{cases}
\frac{1}{4 \gamma \sigma^2} \tonde{1- \sqrt{1- 4 \gamma \sigma^2 \phi}}+ \phi \log \tonde{1+ \sqrt{1- 4 \gamma \sigma^2 \phi}}- \phi \tonde{\frac{1}{2}+\log 2} &\quad \phi \leq  \phi_{\rm May}=\frac{1}{\sigma^2 (1+ \gamma)^2}\\
\frac{1}{2 \sigma^2} \frac{1}{1+\gamma} -\frac{\phi}{2}+\frac{\phi}{2} \log (\sigma^2 \phi) &\quad \phi> \phi_{\rm May}=\frac{1}{\sigma^2 (1+ \gamma)^2}.
     \end{cases}
   \end{split}
\end{equation}
As expected, the functional \eqref{eq:ActionRSn1} does not depend on $q_0, \xi_0, z$ and on the associated conjugate parameters, that have a meaning only whenever more than one replica is present ($n >1$). We consider now the case $n \to 0$, relevant to determine the quenched complexity. It can be shown that $\mathcal{A}_n({\bf x} , \hat{\bf x}, \phi)$ 
admits the expansion: 
\begin{equation}\label{eq:LinearExpansion}
\mathcal{A}_n({\bf x} , \hat{\bf x}, \phi)= n\,  \bar{\mathcal{A}}({\bf x} , \hat{\bf x}, \phi) + o(n).
\end{equation}
Explicitly, for general $\gamma$ we find:
\begin{equation}\label{eq:ActionRSn0}
\begin{split}
\bar{\mathcal{A}}({\bf x}, \hat{\bf x}, \phi)=\bar{\mathcal{p}}({\bf x})+ \mathcal{d}(\phi)+ \hat{q}_1 q_1+\hat{\xi}_1 \xi_1+ \hat m m + \hat p p + \hat \phi \phi- \frac{1}{2} \tonde{\hat{q}_0 q_0+\hat{\xi}_0 \xi_0}-\hat{z} z + \bar{\mathcal{J}} ( \hat{\bf x}),
\end{split}
\end{equation}
where $\mathcal{d}(\phi)$ is as above, while 
\begin{equation}
\begin{split}
\bar{ \mathcal{p}}({\bf x})&=\frac{\tonde{\kappa-\mu  m}}{\sigma^2 (1+\gamma)} \frac{m (q_1-q_0+z \gamma)}{(q_1-q_0)^2}+\frac{\tonde{\kappa-\mu  m}}{\sigma^2} \frac{ p }{(q_1-q_0)}- \frac{\gamma}{2 \sigma^2 (1+\gamma)} \frac{ z^2 (q_1+ q_0) }{ (q_1-q_0)^3}- \frac{\xi_1 }{ 2 \sigma^2(q_1-q_0)}\\
&- \frac{q_0 (\xi_0- \xi_1)}{2 \sigma^2 (q_1-q_0)^2}
 -\frac{1}{2 \sigma^2 (1+\gamma)} \quadre{1+ \frac{2 q_0 z}{(q_1-q_0)^2}}-\frac{1}{2 \sigma^2} \frac{\left(\kappa-\mu 
   m\right)^2}{q_1-q_0} -\frac{\log[2 \pi \sigma^2 (q_1-q_0)]}{2}  -\frac{q_0}{2[q_1-q_0]},
   \end{split}
\end{equation}
and where $\bar{\mathcal{J}} ( \hat{\bf x})$ admits the following integral representation:
\begin{equation}\label{eq:VolumeQuenched}
\begin{split}
&\bar{ \mathcal{J}} (\hat{\bf x})=\int \frac{du_1 du_2}{2 \pi \; \sqrt{\hat q_0 \hat \xi_0- \hat z^2}}  \text{exp}\quadre{\frac{\hat{\xi}_0 u_1^2 + \hat{q}_0 u_2^2 - 2 \hat{z} u_1 u_2}{2 (\hat{q}_0 \hat{\xi}_0- \hat{z}^2)} } \times\\
 &\times \log \quadre{e^{- \hat{\phi}} \sqrt{\frac{\pi}{2}}\frac{1}{\sqrt{2 \hat q_1- \hat q_0}} e^{\frac{(u_1-\hat m)^2}{2 (2 \hat q_1-\hat q_0)}} \text{Erfc} \tonde{\frac{\hat m -u_1}{\sqrt{2 (2 \hat q_1 -\hat q_0)}}} + \sqrt{\frac{\pi}{2}}\frac{1}{\sqrt{2 \hat \xi_1- \hat \xi_0}} e^{\frac{(u_2-\hat p)^2}{2 (2 \hat \xi_1-\hat \xi_0)}} \text{Erfc} \tonde{\frac{-[\hat p -u_2]}{\sqrt{2 (2 \hat \xi_1 -\hat \xi_0)}}} },
 \end{split}
\end{equation}
derived under the assumptions:
\begin{equation}\label{eq:Assumptions}
{\begin{split}
2 \hat q_1 - \hat q_0 >0, \quad 2 \hat \xi_1- \hat \xi_0 >0, \quad \hat q_0 <0  \quad \hat \xi_0<0,  \quad \hat q_0 \hat \xi_0- \hat z^2>0.
\end{split}}
\end{equation}

The saddle point equations fixing the values of the order and conjugate parameters can be obtained taking the derivatives of these expressions, as we recall below. Once the saddle point values are determined by solving the appropriate system of equations, plugging the resulting values into $\mathcal{A}_1$ and $\bar{\mathcal{A}}$ one obtaines the expression for the annealed and quenched complexity, respectively. \\

\textbf{The variational problem and the self-consistent equations.} 
Given the explicit form of the functionals $\mathcal{A}_1$ and $\bar{\mathcal{A}}$, the last step to obtain the complexity is to determine the values ${\bf x}_\star, \hat{\bf x}_\star $ of the order and conjugate parameters that solve the stationarity conditions
\begin{equation}
    \frac{\delta \,\bar{\mathcal{A}}({\bf x} , \hat{\bf x}, \phi)}{\delta \,{\bf x} } \Big|_{{\bf x}_\star, \hat{\bf x}_\star }=0= \frac{\delta \,\bar{\mathcal{A}}({\bf x} , \hat{\bf x}, \phi)}{\delta \,\hat {\bf x} } \Big|_{{\bf x}_\star, \hat{\bf x}_\star },
\end{equation}
as well as the values ${\bf x}^{(1)}_\star, \hat{\bf x}_\star^{(1)} $ that optimize  $\mathcal{A}_1$. In the quenched case, taking the variation of $\bar{\mathcal{A}}({\bf x}, \hat{\bf x}, \phi)$ with respect to the $15$ order and conjugate parameters we obtain two sets of equations of the form ${\bf x}= F_1[\hat {\bf x}]$ and $\hat {\bf x}= F_2[ {\bf x}]$, respectively. These equations couple the $7$ order parameters ${\bf x}$ with the $8$ conjugate parameters $\hat {\bf x}$: inverting one of these sets, one can express the order parameters as a function of the conjugate parameters, ${\bf x}= f_3[\hat {\bf x}]$. The latter can then be fixed by solving the set of coupled self-consistent equations $\hat {\bf x}= F_2[ f_3[\hat {\bf x}]]$: once the self-consistent values of the conjugate parameters $\hat {\bf x}$ are found, the order parameters can be determined and the quenched complexity can be obtained computing the action $\bar{\mathcal{A}}$ at the corresponding values of parameters. The annealed calculation is formally analogous. This scheme can be implemented for generic values of $\gamma$. A detailed discussion of the structure of the self-consistent equations and of the strategy to solve them can be found   in~\cite{PaperLungo}.\\

\begin{figure}[ht]
\centering
\includegraphics[width=0.5\linewidth]{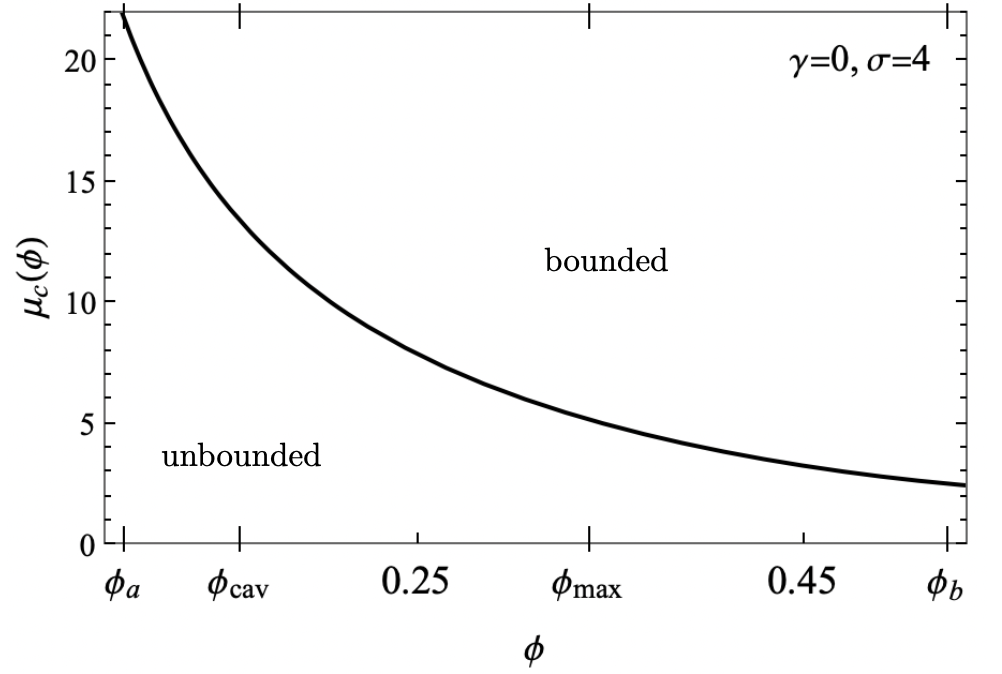}
\caption{Curve separating the unbounded  ($\mu < \mu_c$) from the bounded ($\mu >\mu_c$) phase as a function of the diversity $\phi$.} \label{fig:MuCritico}
\end{figure}

{\bf On the unbounded phase.}  While the quenched complexity $\Sigma(\phi)$ is independent of $\mu$, the typical properties of the equilibria (given by the saddle-point values of the parameters $m, q_1,q_0$) change with $\mu$; in particular, decreasing $\mu$ at fixed $\sigma, \phi$ one finds that the solutions to the self-consistent equations $m^*, q_1^*,q_0^*$ all increase and the system is driven towards the unbounded phase, signalled by a divergence of these parameters
\cite{opper1992phase, GuyCavity, galla2018dynamically, baron2022non}. Given that we have access to the distribution of equilibria as a function of diversity, for each $\sigma$ we can define a $\mu_c(\phi)$ such that for $\mu <\mu_c$ the system is in the unbounded phase. This curves is monotonically decreasing with $\phi$, see Fig. \ref{fig:MuCritico}. This
suggests to define the boundary of the bounded phase in the $\sigma, \mu$ diagram thorough
$\mu^*= \max_{\phi: \Sigma(\phi) \geq 0} \mu_c(\phi)=  \mu_c(\phi_{\rm a}),
$
 to ensure that \emph{none} of the equilibria is in the unbounded phase, no matter their diversity. We remark that the unbounded phase defined in this way has a larger extension with respect to that estimated via the cavity approximation, since 
 $\mu^*>\mu_c(\phi_{\rm cav})$. On the other hand, for $\mu=\mu^*$ the most numerous equilibria having $\phi=\phi_{\rm Max}$ are still in the bounded phase, so the phase boundary obtained using typical equilibria is yet different.
 
\vspace{.3 cm} 

{\bf On the vanishing of the total complexity.}
We claimed in the main text that the total complexity $\Sigma_{\rm tot}=\Sigma(\phi_{\rm max})$  vanishes as $\Sigma_{\rm tot} \sim (\sigma-\sigma_c)^2$ as $\sigma \to \sigma_c^+$ for $\gamma=0$, and that we expect this behavior to extend to $\gamma \neq 0$ \emph{provided that} the maximum of $\Sigma(\phi)$ in the vicinity of $\sigma_c$ lies in a region of $\phi$ in which the annealed calculation is correct. On the other hand, if at the maximum of $\Sigma(\phi)$ the quenched formalism has to be employed, we have indications of the fact that the exponent controlling the vanishing of the complexity is a different one. We motivate these claims in this subsection, and refer to Ref. ~\cite{PaperLungo} for the details. The total variation of $\Sigma_{\rm tot}$ with respect to $\sigma$ is given by:
\begin{equation}\label{eq:TotVar}
\frac{d \Sigma_{\rm tot}}{d \sigma}= \partial_{\sigma} \bar{\mathcal{A}} (\bf{x}, \hat {\bf x}, \phi) \Big|_{{\bf x}^*, \hat {\bf x}^*, \phi_{\rm max}}=\partial_{\sigma} \bar{\mathcal{p}}({\bf x})\Big|_{{\bf x}^*, \hat {\bf x}^*, \phi_{\rm max}}+ \partial_{\sigma} \mathcal{d}(\phi)\Big|_{{\bf x}^*, \hat {\bf x}^*, \phi_{\rm max}},
\end{equation}
where we used the fact that $({\bf x}^*, \hat {\bf x}^*, \phi_{\rm max})$ are a stationary point of $\bar{\mathcal{A}} (\bf{x}, \hat {\bf x}, \phi)$. For $\sigma < \sigma_c=\sqrt{2}(1+ \gamma)^{-1}$, the system is in the unique equilibrium phase and a single, stable equilibrium exists. Its properties (described by the order parameters $m, q_1$) can be derived using the cavity method. For general $\gamma$ and $\kappa=1$, one finds \cite{PaperLungo} that at $\sigma_c$ the equilibrium satisfies
$m=\mu^{-1}=-(1+\gamma)p$, $q_1=(1+\gamma)^2 \xi_1$ and $\phi=\phi_{\rm max}=\phi_{\rm May}= [\sigma(1+\gamma)]^{-2}$. This implies:
\begin{equation}\label{eq:Const}
\partial_\sigma  \mathcal{d}(\phi) \Big|_{\sigma_c, \phi_{\rm max}}= - \frac{\gamma(1+\gamma)}{2 \sqrt{2}}.
  \end{equation}
In order for the complexity to vanish quadratically at $\sigma_c$, this term should be compensated by the one obtained deriving  the distribution of the forces $\bar{\mathcal{p}}({\bf x})$. If for $\sigma>\sigma_c$ and $\phi=\phi_{\rm max}$ the annealed calculation is exact, than one can replace $\bar{\mathcal{p}}({\bf x}) \to \mathcal{p}_1({\bf x})$, and use that for the values of parameters predicted by the cavity approximation it holds:
\begin{equation}
\partial_\sigma \mathcal{p}_1\Big|_{\sigma_c, \phi_{\rm max}}= \frac{\gamma (1+ \gamma)}{2 \sqrt{2}},
\end{equation}
which cancels exactly \eqref{eq:Const}. Therefore, if $\Sigma_{\rm tot}$ is analytic at $\sigma_c$, it has to vanish quadratically (one can check that the second derivative is not vanishing at the critical point). On the other hand, for $\gamma=0$ we know that at $\phi_{\rm max}$ the annealed calculation is never correct, for any $\sigma>\sigma_c$. Assuming that this is still true for $\gamma=0$, imposing that \eqref{eq:TotVar} vanishes and using the conditions given by the cavity approximation (in addition to $q_0=(1+\gamma)^2 \xi_0$ by symmetry) we obtain the following conditions for the order parameters:
\begin{equation}\label{eq:Condition}
 \frac{ z}{(1+\gamma)(q_1-q_0)^2} \tonde{\frac{\gamma z (q_1+q_0)}{2 (q_1-q_0)} + q_0}=0,
\end{equation}
which implies either $z=0$, or $z=2 q_0 (q_1-q_0)/[\gamma (q_1+q_0)]$. Both these solutions however can be shown to be incompatible with the quenched self-consistent equations for this order parameter~\cite{PaperLungo}  except for the case $\gamma=0$, when in fact it holds $z=0$ at the transition point. Therefore, if for $\gamma \neq 0$ the total complexity at $\sigma \sim \sigma_c^+$ is quenched, one should expect a different power law since the linear contribution is not vanishing.   
We remark that the symmetric case $\gamma=1$ is special, since the total complexity should vanish in a non-analytic way at the transition, due to the square root term in \eqref{eq:DetSmall22} whose argument vanishes when $\phi=\phi_{\rm May}, \sigma=\sigma_c$.

\end{document}